# LLM tools in the prediction of the stability of perovskite solar cells

S. Frenkel[1], V. Zakharov[1], E. A. Katz[2],*

[1] Federal Research Center "Computer Science and Control" Russian Acad.of Sc., Moscow, Russia
[2] Ben-Gurion National Solar Energy Center, Swiss Institute for Dryland Environmental and Energy Research, Jacob Blaustein Institutes for Desert Research, Ben-Gurion University of the Negev, Midreshet Ben-Gurion, Israel
*e-mail: keugene@bgu.ac.il

**Abstract**

Predicting degradation rates is an important task in the development of new perovskite solar cells (PSCs). In this paper, we explore the feasibility of solving this problem using Machine Learning models supported by LLM tools.

We consider both the "lifetime" prediction of the device and the prediction of degree of its degradation at specific time intervals. We demonstrate the ability of common LLM tools (ChatGPT, DeepSeek) to suggest and justify prediction methods in a dialogue with the user under conditions of incomplete information about the physical models of PSC degradation and the influence of the environment, providing rather accurate prediction. The results cover the formation of time series of efficiency with a given architecture, calculated using various available mathematical models together with environmental characteristics archived in various meteorological databases (illumination, temperature, humidity, UV level). We conclude that ChatGPT currently has sufficient access to training samples, can suggest to PSC designer various PSC models in the literature, and has adequate solutions for predicting degradation trends.

## 1.Introduction

Currently, the task of evaluating the stability of solar energy conversion into electrical energy by perovskite solar cells (PSC) is of great interest.

PSCs are a new type of solar cell that provides a fairly high-power conversion efficiency (PCE) compared to silicon, lower production costs, and minimal energy consumption during production [1]. However, PSCs suffer from stability problems, which are understood as a decrease in energy conversion efficiency during operation to unacceptable values, for example, to 80% of the initial value, and in this case, the lifetime (LT) is denoted as T80, although other thresholds are also widely used, for example, 50% (T50, respectively) [2]. The enormous amount of research and measurements required to ensure stability, consisting of testing a significant number of variants of the materials and measurement methods used [3], creates a need for the use of modern ML tools to solve the problem of predicting LT, including Large Language Models (LLMs). At the same time, due to the short history of research and development compared to the required lifetime of PSCs (< 25 years), there is no representative statistics for statistical estimates of LT that could be used in ML prediction algorithms. The application of accelerated testing (AT) techniques [2], while a desirable solution, is often difficult in practice due to the enormous number of factors influencing PCE (weather, technological, solar spectrum, etc.) and the impossibility of realistically reproducing them in laboratory (indoor) conditions, often due to the high cost of laboratory equipment [3]. The uncertainty in knowledge about the degradation mechanisms of PSCs prevents reliable extrapolation of results to the distant future based on AT results. Therefore, methods of prediction based on field (outdoor) measurements, supplemented, where possible, by data from other sources [4], are of primary interest.

In our opinion, one of the specific aspects of modern PSC design is that in studies on predicting PSC LT using ML methods, along with experimental data from specific PSCs, developers are forced to use data extracted from various databases of published articles [5] in order to somehow overcome the problem of obtaining more reliable statistics on admittedly insufficient numbers of samples, thus trying to increase the sample size.

In this paper, we investigate whether LLM tools, under the conditions described above, can provide new solutions to the PSC developer. By "novelty," we mean obtaining solutions ("suggestions") that are explicitly or implicitly absent from the queries to the LLM.

Recently (2025), studies [6] have emerged describing the development of multi-agent LLM systems for designing PSCs, with a well-thought-out question-and-answer system, and which, according to the data presented, answer the developer's questions significantly more completely than ChatGPT. However, they are focused on materials science aspects, not including in the subject area and its ontology such important components of knowledge as physically based models of PCE degradation and ML models, and require the inclusion of qualified programmers in the teams to develop such LLM agents. At the same time, when comparing with ChatGPT, the non-determinism of the answers is significant, which will be discussed in Section 2.

In contrast to this, we demonstrate that state-of-the-art LLM models can solve complex prediction problems without the need for multi-agent LLM submodels to solve specific subproblems of the prediction task.

That is ChatGPT is capable of independently identifying subproblems that require a specific solution. This is because the available ChatGPT ontology allows the formation of time series of PCE values for a given PSC architecture, calculated using various available models, which, together with environmental characteristics archived in various meteorological databases (illumination, temperature, humidity, UV level) for the desired precise time intervals within a given season, for a given location (e.g.,Berlin, Beer-Sheva). The obtained time series can be considered as synthetic data, and they can be used together with measured data to predict LT.

Of course, ChatGPT allows us to upload real measurement data, for example, as a CSV file, but the specifics of the measurements require certain efforts to compare the results of different laboratories (more on this in Section 3). Therefore, the use of synthetic degradation trajectories, which maximally consider the specifics of a particular architecture and environmental parameters while controlling the correspondence of individual outdoor measurements, is the optimal choice for studying the effectiveness of LLM in solving the indicated problem. It should be noted that this approach has been used for decades to test the performance and reliability in the development of complex computing systems, using various synthetic benchmarks.

The acceptable accuracy of T80 prediction is demonstrated both over short intervals (1 month) and over a year, showing that the degradation models selected by the LLM correctly convey the direction of the degradation trend.

Important information for readers may be an example of a detailed description of the search and use of data from various databases and published sources to build the models used in the prediction.

Let's make a few clarifications about how we will present the texts generated by ChatGPT:

- Texts generated by ChatGPT, transmitted without changes, are enclosed in quotation marks, while any text with the personal pronoun "I" was generated by ChatGPT.

- Texts not enclosed in quotation marks are our interpretation of the ChatGPT texts.

Let us also clarify that "user" refers to the PSC researcher who is in a dialogue with ChatGPT in this chat session.

## 2. LLM in Solving the Problem of Predicting PCE Degradation

Let's introduce some necessary concepts from LLM and PSC-area.

### 2.1. Subject Domain

The most important concept used in the practice of applying LLM is *subject domain of a problem*. In modern practice of developing software and information systems, the subject domain (SD) is a segment of representations about reality (for example, physical phenomena, measuring instruments, measurement methods), for processing data about which a specific software and hardware system must be developed, capable of receiving and processing data from this "reality," presented in documentation, publications, or practical activities (in the form of experience (documented or not) of development participants), in terms of relevant theories and technologies [6].

### 2.2 Non-determinism of LLM Answers

The phenomenon of non-determinism of ChatGPT and DeepSeek answers is well known, where completely different answers are unpredictably given for the same query [7]. The main reason for ChatGPT's non-determinism, which The ChatGPT names “Randomness Generation" is: “even if the same model is used with identical formulations, the answer will differ depending on the so-called “temperature settings”. “Temperature” is a parameter belonging [0,1] that controls the probability of word selection. A higher “temperature” (e.g., 0.7) leads to a more diverse set of ontology elements, while a lower one (e.g., 0.2) makes the output more deterministic.

However, as our experience shows, this can also have a positive effect, since non-determinism leads to the reflection of different aspects of the problem to which the questions relate in the answers, and their combination can help to solve the problem more completely. This will be discussed and shown in Subsection 2.4.

### 2.3 PSC characteristics used in PCE degradation prediction models

For a more detailed further explanation, let's give a brief description of the main concepts of the PCS LT prediction problem.

1) **Architecture**. The type of PSC architecture, according to [10], is determined by a layered structure consisting of a transparent conductive oxide layer (TCO), an electron transport layer (ETL), a perovskite light-absorbing layer, a hole transport layer (HTL), and a back electrode. Example architecture: FTO/CB-TiO2/MP-TiO2/Triple Cation/Spiro-OMeTAD/Au, where FTO (fluorine-doped tin oxide) as the conductive substrate, c-TiO2 (compact layer of titanium dioxide), mp-TiO2 (mesoporous layer of titanium dioxide). The structure of the PSC consists of a transparent conductive oxide (TCO) layer (e.g., indium tin oxide or titanium dioxide), a triple-cation perovskite absorber layer, Spiro-OMeTAD (a hole-transporting material), and finally, a metallic gold (Au) contact layer at the top.

2) Target electrical characteristics that characterize the ability to convert solar energy into electrical energy:

- Power Conversion Efficiency, PCE(t) = E(Pout(t))/E(Pin(t)) [2, 3],

where E(Pout(t)) is the average power of electrical energy generated by the PSC, E(Pin) is the average power of solar energy, t is the current time over the averaging interval,

- $J_{sc}$ - the current at which the potential difference across the solar cell is zero,

- $V_{oc}$ - the potential difference at which zero current flows through the device,

- FF - fill factor. This is a measure of the "rectangularity" of the PSC's current-voltage characteristic curve. On the curve, there is a point where the values of J and V are maximum (Imax, Vmax), forming the maximum power point (MP), $FF = J_{max} \times V_{max} / J_{SC} \times V_{OC}$.

3) Environmental parameters (temperature, illumination, humidity, etc.)

4) Physical models describing the behavior of the target electrical characteristics, typically Arrhenius and Eyring models [9], in which the role of stress factors is played by temperature, external electrical voltage, load current, etc.

5) Mathematical and software prediction tools (Prediction Models) - mathematical models of PCE efficiency degradation, mathematical models of the influence of external environmental factors, statistical models of the dispersion of characteristics of PSC samples, types of ML-tools (various types of Neural Networks, gradient-based tools, regression model-based).

### 2.4 About ontology of the problem subject domain

An ontology informally represents the domain of knowledge by defining the key entities, their properties, and the relationships between them. Ontologies are used in software tools to provide a structured, semantic model of concepts, relationships, and data within a domain.

In the most general sense, the completeness of answers to questions posed by the ChatGPT is determined by the degree of overlap between the ontologies of the PSC developer's subject domain of the PSC lifetime prediction problem and the subject domain visible ChatGPT. Essentially, it shows the concepts and their relationships that ChatGPT can process to answer the input questions and perform searches in relevant publications and PSC databases (e.g., [5]).

The user can use these same concepts from subject domain ontology and relationships for queries.

In fact, the ontology provides a checklist of entities to consider. If key entities are missing from the forecasting methodology, this indicates the impossibility of obtaining a component of the solution corresponding to the missing entity, or the relationships between entities.

Note that the same entity can be called by different terms or expressed mathematically (also in different terms).

For example, the entity "non-monotonic degradation of the PCE curve" is equivalent (given the same environmental behavior (illumination, temperature, humidity, power of ultraviolet radiation (UV))):

- to the presence of self-recover / Reversible Degradation Mechanism properties [4] (the entity "Reversible Degradation Mechanism"),
- to the Ornstein-Uhlenbeck (OU) stochastic process model – the entity "OU", since from the point of view of the physicochemical processes occurring in the PSC, degradation is represented by the Reversible degradation equation (that is, a non-monotonic solution) [16]).

If the ontology does not contain a type of PCE behavior characterized (directly or indirectly) by "non-monotonicity," the forecast will fundamentally be incomplete. For example, it may only predict degradation due to thermal and light exposure, ignoring internal causes of degradation (e.g., ion migration).

We will not provide the generated complete description of the ontology concerning material properties, architecture, environment, and testing methods. The production process, totaling over 1100 words (excluding detailed comments), will only show the components of the ontology related to degradation processes, since these (with the given characteristics related to the aforementioned ontological components) determine the numerical characteristics of stability, and lifetime in particular.

**Query to Chat GPT**:

*build Ontology of the Subject Domain for Outdoor-Based Lifetime Prediction for Perovskite Solar Cells considering all PCE degradation aspects including recovering of degradation*

Answers related to the degradation process:

"…. entities and relations to capture dynamic degradation and recovery cycles":

**DegradationPhase:** A contiguous time interval showing monotonic efficiency loss.
**RecoveryPhase:** A time interval showing partial PCE restoration.
**Subclass of DegradationPhase or peer class linked via alternates With**:
ReversibleMechanism Subclass of DegradationMechanism for processes that can self-recover (ion migration, reversible defect passivation, etc.)
**IrreversibleMechanism:** Subclass of DegradationMechanism for permanent damage (chemical decomposition, delamination, corrosion).
**HysteresisEffect:** Optional subclass capturing IV curve hysteresis and short-term recovery within measurement cycles.

**Key object properties**

**hasDegradationPhase Device → DegradationPhase Device's:** full degradation timeline segmented into phases.
**Alternates with DegradationPhase ↔ RecoveryPhase**: Indicates cyclic alternation.
**causedBy RecoveryPhase → ReversibleMechanism:** Links observed recovery to physical cause.
**hasReversibilityRatio (data) DegradationPhase**: Ratio of recovered PCE to lost PCE
**recoveryTimeConstant_h (data) RecoveryPhase**: Time constant for recovery (for kinetic modeling).
**recoveryCondition (data):** RecoveryPhase Qualitative description ("dark rest", "no bias", "nighttime cooling").

**Feature Engineering** (using all the listed data + Performance Metric),
**- Data preparation:** time-series light exposure, thermal stress history,
**- Model selection -** types of models and their properties, such as:
**-** Random Forest (RF) **-** tolerant to measurement noise, sensitive to non-linearity of the relationship between variables and response, but the results are poorly interpretable,
**-** Gradient Boosting (XGBoost, LightGBM) **-** accuracy depends on the choice, etc. for others.

The answer also describes the training and model validation process (model validation and testing), the Hyperparameter tuning scheme, for example, using the GridSearchCV function (from Python libraries) for iterating through hyperparameters, Cross-validation (k-fold, time-series split), Feature Importance Analysis (SHAP, LIME methods [16]) to identify key degradation factors,
**selection of performance metrics for the prediction tool (model) (Performance Metrics:** Root Mean Squared Error (RMSE), Coefficient of Determination, Goodness-of-fit Mean Absolute Error (MAE) .
**Model training**: Include recovery descriptors as features to avoid over-penalizing reversible degradation when predicting long-term T80.

Thus, the presented ontology allows us to hope that when requests solving problems related to calculating certain stability characteristics, such as a particular characteristic of the degradation rate, ChatGPT will search for the answer among publications and databases containing the specified knowledge. For this, the queries must explicitly refer to the generated components of the LT prediction problem ontology.
Note that in the case of "temperature" =0 (see Subsection 2.2), the ontology of the subject domain for predicting the service life of perovskite solar cells (PSC) for indoor and outdoor applications, the LLM selects only the most probable words when generating the ontology text, which means using only accurate, non-controversial knowledge. However, as noted in Introduction, a few degradation mechanisms are currently controversial among PSC specialists. The high degree of determinism of the answers ("temperature" near zero) reduces the likelihood of accounting for the lack of consensus among specialists regarding the validity of certain physical models, which, as noted in the introduction, is an important factor influencing PSC design work. The ability to account for this circumstance when selecting and using prediction tools is highly valuable. For example, in response to the query "uncertainty and ongoing debates around the PSCs stability mechanisms," GPT and Deepeek provide the answer: Ion Migration and Its Long-Term Effects, but they also provide nine discussion topics and a list of references relevant to the still-debatable ontological issues regarding the nature and extent of the influence of Ion Migration and its interaction with the ontological entities of DegradationMechanism: MoistureInduced, ThermalStress, LightInduced, IonMigration, OxygenDegradation, etc.

Because these stability aspects are highly debated issues, their models are relatively rarely mentioned in publications on practical stability prediction, and the likelihood of their presence in LLM responses is low, as we will see in further analysis of ChatGPT responses. This suggests that before selecting a model, it is important to find out (from GPT, DeepSeek, etc.) the degree of study and debate surrounding the models in a given problem, and, if necessary, include a proposal to use an appropriate degradation model in the query.

## 3. Predicting the Degradation Process of PSCs

### 3.1 PCE Measurement Practices

Let's briefly consider the practice of outdoor (field) measurements at relatively early stages of PSC development.

Currently, the main measuring instrument for studying PSCs are multi-channel setups, to which dozens (sometimes up to hundreds) of PSC elements can be simultaneously connected, measuring their maximum power points (MPP) [8], which involves tracking and maintaining the voltage and current that provide the maximum possible output power of the element. Since PSCs often exhibit strong hysteresis, which makes it difficult to accurately determine the true maximum power point using standard tracking algorithms, this requires very expensive equipment (Fig. 1) [8]. The need for simultaneous connection of many (up to hundreds) of the studied PCS elements is due to the long measurement times (months, or even years - see below), and therefore it is desirable to observe as many elements as possible simultaneously, which requires a variety of tests, usually performed by different researchers. Given that the channels themselves (slots, with the necessary matching electronic components [8]) are also expensive and purchased separately, the question of their rational use is relevant. Replacing elements for which the forecast is unsatisfactory at some point in time, for example, approaching the threshold of permissible degradation, counting the time from the beginning of degradation ("aging").

### 3.2 An Example of ChatGPT-based prediction

A recently published paper [11] for the first time presents observations of PSC degradation over four years for two PSC samples measured in Berlin in 2021-2024 (reproduced in Fig. 1 (captions under the figure from [11] are slightly shortened)) with the architecture:

ITO | 2PACz | $Cs_{0.15}FA_{0.85}PbI_{2.55}Br_{0.45}$ | $C_{60}$ | $SnO_2$ | Cu

It should be noted that such a sharp drop in performance during the autumn-winter period, despite subsequent recovery, may be unacceptable, because with such low efficiency, the PSC element may not have enough power to solve a specific task during the specified period. Therefore, if this aspect is considered important, it may be useful to predict degradation to some unacceptable level, either to deem this PSC element unpromising for further research, or, if it is an industrial sample, to replace its element for the winter period. The similarity of this task to the task of early rejection is obvious, in the sense that its goal is to predict the local lifespan, i.e., the time of acceptable degradation in a certain time interval of interest to the researcher at a considered moment.

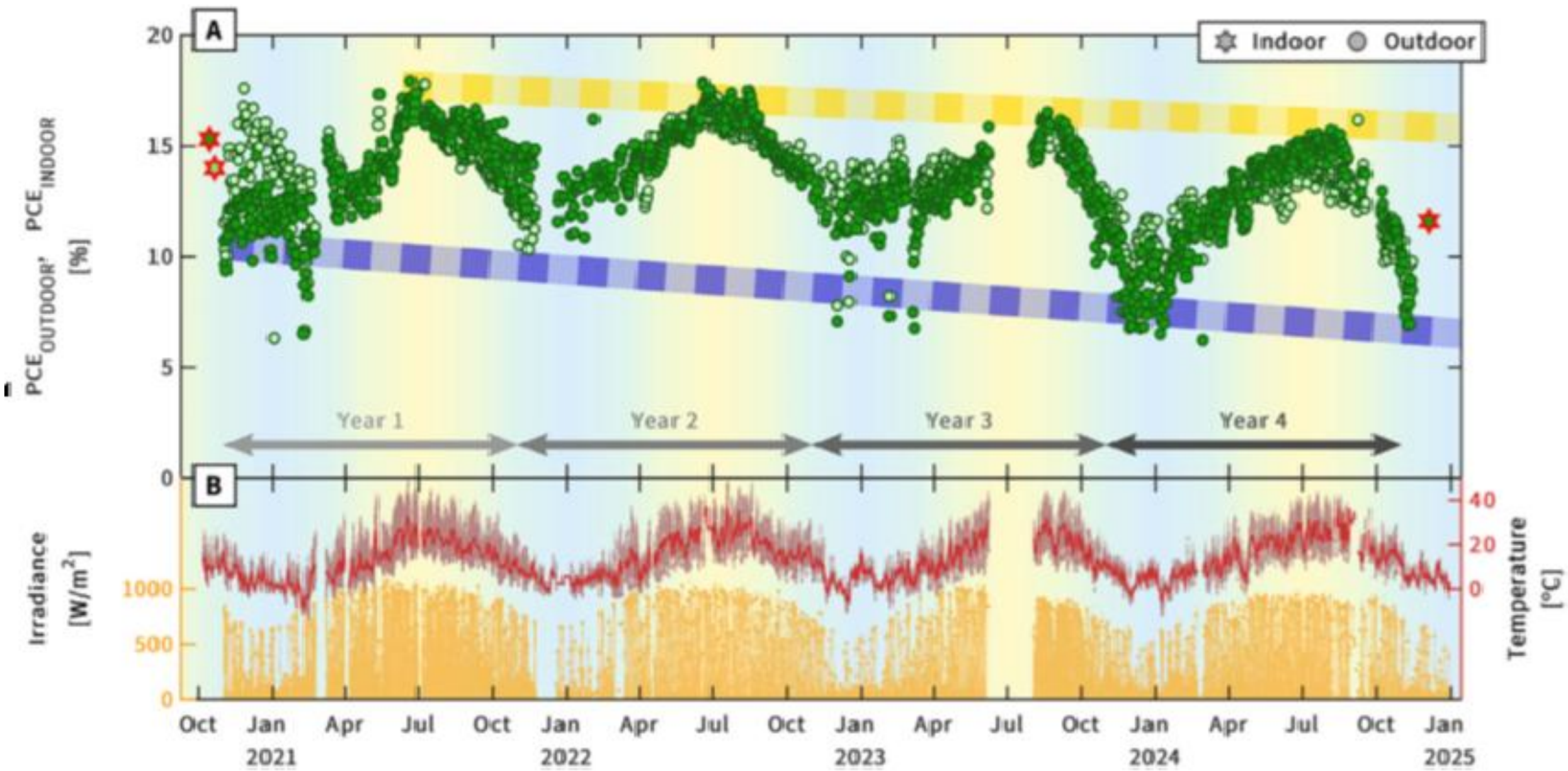

Figure 1. A) A Long-term outdoor PCE [11]. The yellow and blue dashed lines represent hypothetical linear degradation of 3% and 9% per year, respectively, serving as a benchmark for visual assessment of PCE OUTDOOR changes during the summer and winter periods. B) Temperature measured on the back of the device's encapsulation layer and irradiance in the plane of the array. The background color gradient (in all graphs showing outdoor measurement data in this paper) represents the change of seasons – winter is indicated in blue, summer in yellow. Reproduced from Ref. 11 with permission.

We formulate a request to ChatGPT:

*daily prediction for September-November of 2023 PCE element*
ITO $|2PACz|Cs_{0.15}FA_{0.85}PbI_{2.55}Br_{0.45}|C_{60}|SnO_2|Cu$, *by available the PCE, irradiance and time series for 2023 in Berlin, using both XGBoost regression and hybrid model XGB+OU, with 6 first months data of 2023 for training, 2 months for testing.*

(OU stands for the Ornstein-Uhlenbeck stochastic process, which will be discussed below).

ChatGPT reports that some information about PSCs *with this architecture is mentioned in the paper [11]*, and asks whether to use the data from this paper or whether the user will provide their own.
After the suggestion to use this data to build prediction models, the search begins.
We will present (partially in summary) the ChatGPT responses.
First, ChatGPT states that since the actually measured PCE time series for 2023 are not publicly available for download from the webpage, the presented demonstration of prediction capabilities uses synthetic, *but physically plausible*, PCE time series (ChatGPT provides a detailed description of the data calculation model, taking into account the initial PCE value from [11], clearly documented and easily replaceable (in the generated Python) with real data).
These actions obviously reflect the specificity of PSC element development mentioned in the introduction, which is due to the need to combine proprietary laboratory data and the use of data from other laboratories.

Since the degradation process depends on daily changes in illumination and temperature, ChatGPT reports that it will search for:
“Dataset: synthetic Berlin 2023 daily irradiance, temperature, used to build a real synthetic daily PCE time series”.

Further, ChatGPT suggests and explains the method for obtaining synthetic data:
It searched databases and found the data required for prediction described in the paper [12] indicating that the paper “available on reasonable request from the corresponding author”.
To obtain the necessary time series (hourly) data for predicting solar power and air temperature in Berlin for the time interval under consideration, ChatGPT found and agreed with the user on the databases (ChatGPT's suggestion and the user's response in the same chat), stating that it would search for daily PCE predictions (meaning that the averaging of absorbed solar radiation power and generated electrical power (see 2.3) occurs over a 24-hour period) in September-November 2023 (forming and storing the "synthetic Berlin dataset"; note that a significant portion of current publications consider precisely the daily (24-hour) average).The following databases were named DWD (German Weather Service) Climate Data Center — station observations (10-min, hourly, etc.), and ERA5/ERA5-Land (Copernicus / ECMWF— hourly gridded reanalysis).
Considering the PSC operation models used in the prediction, ChatGPT also searched for global horizontal irradiance in SOLARGIS / Global Solar Atlas / satellite datasets, and reports that:

"PVGIS / JRC provides hourly/daily irradiance data and can be used as the irradiance source for Berlin 2023 if needed. Use PVGIS hourly or the PVGIS REST API to download hourly irradiance for coordinates (52.52 N, 13.405 E)".
(PVGIS is a free web application that allows the user to obtain data on solar radiation and energy production from photovoltaic systems).

In addition to the resources listed, ChatGPT reports finding the Zenodo PSC Perovskite ageing MPPT dataset (Zenodo: Perovskite Solar Cells Ageing Dataset [13]). This dataset was used in an analysis published in Nature Communications in 2023 and contains graphs of MPPT efficiency versus time for multiple elements (pickle .pkl file). In it, ChatGPT identified the requested PSC (ITO|$2PAC_z$ | $Cs_{0.15}FA_{0.85}PbI_{2.55}Br_{0.45}$ |$C_{60}$|$SnO_2$|Cu) (an instance of the triple cation group), presented in the Zenodo dataset. The Zenodo dataset contains metadata fields listing the absorber composition and layer materials.
Importantly, this database is not referenced in the paper under consideration, and as we found out, it is supplementary material to an earlier paper by the authors [12].
This is an important circumstance, in our opinion, demonstrating a broad coverage of the subject area, going beyond simply viewing and structuring publications, as LLM tools are often characterized.
ChatGPT provides a detailed description of the use of XGB and Hybrid XGB+OU prediction models, which is not included here. The response to the request addressed to the specified ChatGPT datasets is accompanied by the following message:
The result of ChatGPT's work on the given request is:

- Python code of the prediction and results analysis model, which includes a "config" section that specifies access to all external databases (some of which will be discussed below) containing information, and which, accordingly, can be adjusted depending on the architectures, data, etc. used,
- curves of the modeled PCE time series (named "actual" by ChatGPT, since all real PSC parameters of the specified architecture taken from the verified database are used for modeling), predicted using XGB, and a hybrid XGB+OU model (Fig. 2),
- evaluation of the values of the main prediction quality metrics - a detailed analysis and explanation of the chosen model for synthesizing "actual" series, and a comparative analysis of XGB Regression and XGB+OU, with an analysis of the scheme (parameter selection) of their use.

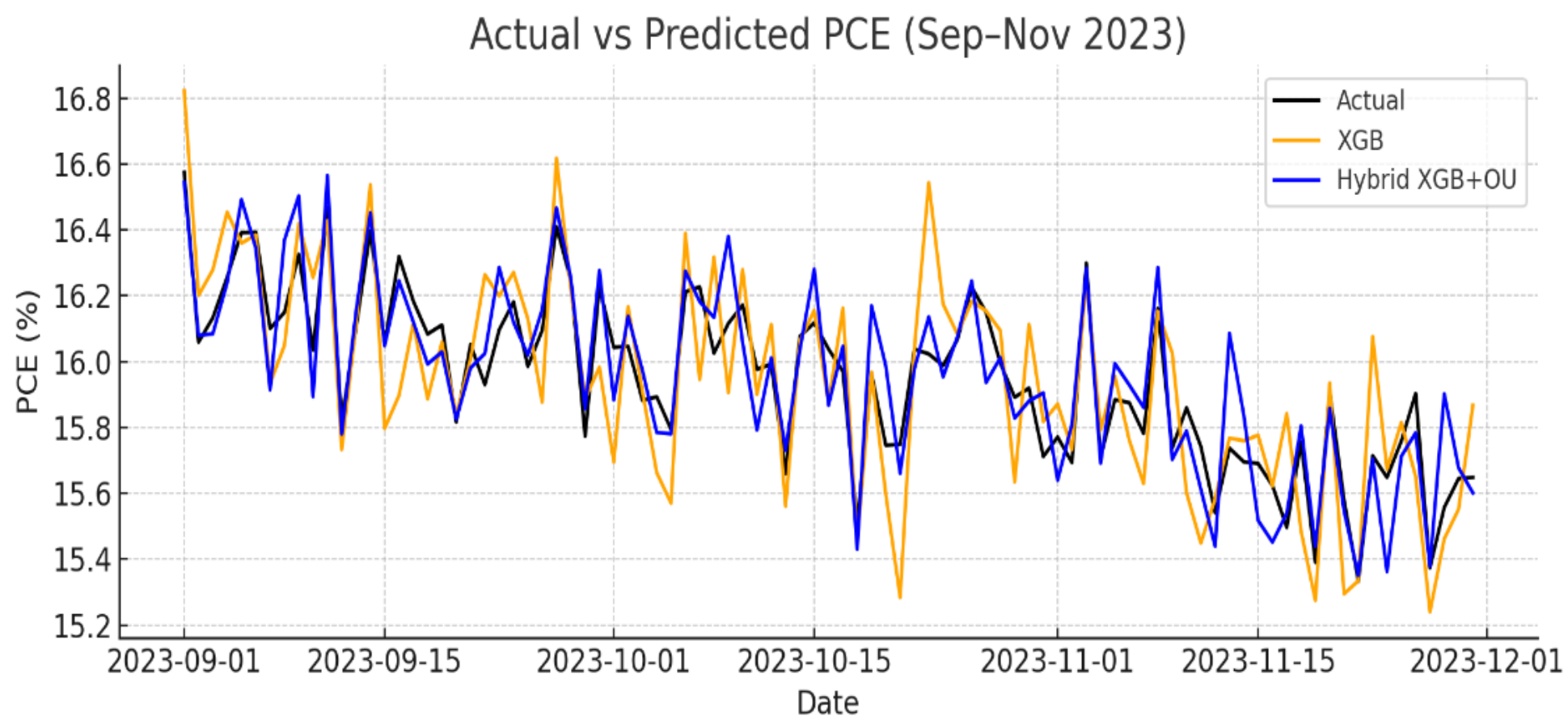

Figure 2. Actual and predicted degradation curves of the PSC for the architecture under consideration.

Note that it is easy to estimate the corresponding T80 value as the maximum (rightmost) root of the equation y(t)/max(y(t))=0.8, where y(t) is the corresponding PCE time series.

We will not provide the prediction quality values calculated and displayed by ChatGPT, we will only say that, as can be seen from the Figure, the maximum relative prediction error is < 5%, and the correlation coefficient $R^2$ = 0.95-0.98 for the test period (June-August), and 0.86-0.85 for the predicted period), which indicates a high agreement between the predicted and actual data), and we note that the predicted curves completely repeat the downward trend of the measured data in the predicted interval.

Correct prediction of the trend direction, as mentioned above, we consider as the main criterion for the quality of the prediction, given the indicated problems with the accuracy of MPP measurement and the precise knowledge of some degradation mechanisms.

ChatGPT indicates in its answer to the formulated prediction task:

"Absolute PCE values will vary from sample to sample and depending on MPP measurements. Comparative indicators presented in the literature [13] show that the nature of PCE degradation varies greatly — the expected standard deviation of daily PCE will be in the range of several tenths of a percent of the absolute PCE."

Therefore, some difference in PCE values in the published graphs and synthesized (modeled) time series is not fundamentally important. It is important that ChatGPT generated a well-structured text to explain the method of finding and using the data: “The long-term HZB (Helmholtz-Zentrum Berlin) dataset for outdoor conditions provides useful real-world trends (seasonality, stratification events), but if you need formal T80 values according to specific ISO protocols (e.g., damp heat, 85°C/85%RH, MPP tracking), we may need to extract data from individual stability studies or request complete datasets.”

Note that this remark refers to the situation where, due to the small number of elements, it is possible to try to use prediction results from the database for the same material and environmental characteristics for forecasting (see Section Discussion and Conclusion for details).

Unlike the frequently considered task of managing the testing time of a specific PSC element, in the case currently under consideration, ChatGPT correctly takes into account that the general task of estimating the time of permissible degradation (T80) is *to answer the question about T80 not for an individual element, but for a certain batch of elements based on a sample of a certain number of tested elements* (a typical situation of sampling inspection). ChatGPT provides the following solution to this problem:

"When multiple matching devices are available, the Notebook Python offers three options (select one in the code or via an argument):

aggregate by median PCE for matched devices -> one target,

select the most suitable single device based on metadata, train separate models for each device and an ensemble."

**3.3 Description and analysis of the mathematical models used by ChatGPT for PCE time series generation**

Since the generation of the predicted PCE time series occurs through calculations using physics-mathematical models of electrical energy generation with PSCs, and as noted in Section 2, the physics-chemical mechanisms of this generation remain debatable, it is interesting to see how LLM tools select these models.

To verify and analyze the models used to obtain the actual PCE, we formulated the following query (in the same chat):

*describe the way that you used to build actual PCE time series. Did you use mathematical models? Describe them if so.*

ChatGPT responded to this query with the following answer and analysis:

"Yes, I did use a mathematical, physics-inspired model to generate the synthetic "actual" PCE time series for the Berlin 2023 dataset. Because we don't have measured values here, I built a model that mimics the real physical behavior of a perovskite solar cell (PSC) ITO |$2PACz$|$Cs_{0.15}FA_{0.85}PbI_{2.55}Br_{0.45}$|$C_{60}$|$SnO_2$|Cu.

"Short description of the synthetic PCE model I used (so you can judge/adjust)

A physical–empirical equation for "true" PCE was used, modeling the (quasi-deterministic) trend, and an Ornstein–Uhlenbeck (OU) latent "device-state" variable captures slow metastability / aging / light-soaking drift" (the random component of the trend).

1. The physical–empirical model is a generalization of the Eyring model [9], used to model the influence of various factors on the operation of semiconductor devices. The model is a product of exponentials, each factor of which corresponds to one of the known photovoltaic, temperature, and spectral mechanisms, for example, with the components:

   - $(1-\beta(T-25))$ - Standard negative temperature coefficient for perovskites (~−0.2 % / K) ("−0.2 %/K" for perovskite means that the material property changes by −0.2% for each increase in temperature by 1 Kelvin (K)),
   - $(1+\gamma(A-1.9))$ - Slight optical modulation by spectral air-mass and path length.

ChatGPT provides a complete view of the model with a detailed description of all parameters, so that the user in the current chat can clarify both the theoretical foundations of such use and the numerical values calculated from "historical" data from the relevant datasets.

Random fluctuations in the ChatGPT degradation trend were modeled as an OU process (Section 3.3), explaining that:

“The OU process is a stochastic differential equation defined as:

$dS(t) = -\theta S(t)dt + \sigma dW(t)$

• S(t): The short-term deviation from the long-term trend.

• θ (theta): Mean reversion rate. How quickly the fluctuations revert to zero.

• σ (sigma): Volatility. The magnitude of the random fluctuations.

• dW(t): The increment of a Wiener process (Brownian motion), representing random noise”.

OU drift gives low-frequency correlation, matching real perovskite stability experiments”. The characteristics/data used for synthetic PCEs are:

G(t) - Irradiance Annual pattern with daily noise - Berlin-like daylight cycle at time t,

T(t) - ambient air temperature data.

It is reported that the OU drift parameter is calculated from the dataset used in the analysis published in the paper [13], and containing graphs of the dependence of PCE (estimated from measured MPP) on time for many samples, and that the light-induced metastability model takes into account past integrated illumination using an exponential kernel.

The main parameters of the models were formed from the empirical data presented in the paper under consideration, and ChatGPT calculated them without user requests.

At the same time, ChatGPT explains that the phenomenon of Ion Migration in PSCs is:

“Modeled as a random walk or diffusion process with a known probability distribution for the rate of ion movement)”.

ChatGPT reports that thanks to this combination of models, it is possible to obtain realistic seasonal, daily, and slow aging patterns on the PCE curve because:

“This hybrid of physics and statistics (Ornstein–Uhlenbeck (OU) models.

Physical realism → temperature and irradiance have correct signs/magnitudes,

Machine-learning readiness → outputs can be used for training regressors like XGBoost or hybrid XGB+OU,

Ornstein–Uhlenbeck (OU) models captures slow metastability / aging / light-soaking drift) allows:

Reproducibility → parameters can be tuned to match real measured Berlin data once available”.

ChatGPT reports that the two models provided are combined as follows:

$PCE(t)=K(1+\alpha \ln(G(t)+1))(1-\beta(T(t)-25))(1+\gamma(At-1.9))+\delta O(t)+\varepsilon_t$

with the constants used in the synthetic generator:

$K=15.3\times 0.88$ (in percent),

$\alpha=0.08$ (sensitivity to illumination),

$\beta=0.0025$ (temperature coefficient),

$\gamma=0.5$ (spectral factor),

$\delta=9.5$ (parameter value related to the OU state),

$\varepsilon_t$ is a small noise (set to 0 for sensitivity graphs)".

(Thus, the user has the opportunity to manually check and analyze the operation of ChatGPT, and the opportunity to ask a question in the same chat in case of any discrepancies).

And it reports:

"built a model that mimics the real physical behavior of a perovskite solar cell (PSC) stack like ITO |$2PAC_z$|$Cs_{0.15}FA_{0.85}PbI_{2.55}Br_{0.45}$|$C_{60}$|$SnO_2$|Cu under Berlin-like irradiance and temperature variations ".

Obviously, the important question is what PSC parameters were used when generating the PCE time series.

To the query:

*give the parameters of ITO | 2PACz | $Cs_{0.15}FA_{0.85}PbI_{2.55}Br_{0.45}$ | $C_{60}$ | $SnO_2$ | Cu, you used to sensitize PCE*

ChatGPT respond:

"The goal was just to produce plausible daily PCE fluctuations given environmental conditions (irradiance and temperature). Here's a breakdown of what I assumed for each layer/parameter effect:

"The parameters were found from the tables in the paper [14] irradiance effect: 2tanh(irradiance/5) saturates at high irradiance (~+2 %) by assumption:

| Layer | Role / Effect in model |
|---|---|
| ITO | Transparent electrode – assumed ideal, no loss modeled |
| 2PACz | Hole-selective layer – assumed perfect extraction, no added resistance |
| $Cs_{0.15}FA_{0.85}PbI_{2.55}Br_{0.45}$ | Perovskite absorber – main variable affecting PCE; PCE depended on irradiance (tanh(irradiance/5)) and temperature (-0.05*(T-20)) |
| $C_{60}$ | Electron transport layer – assumed ideal, no series resistance |
| $SnO_2$ | Electron transport/ETL – no added effect beyond baseline PCE |
| Cu | Back contact – assumed perfect, no ohmic loss |

At the same time, ChatGPT noted that "There is not a "universal perovskite database" that contains all device stack parameters (e.g., thicknesses, mobilities, trap densities, interfacial band offsets) for every specific device architecture, especially exotic stacks like ITO | 2PACz | $Cs_{0.15}FA_{0.85}PbI_{2.55}Br_{0.45}$ | $C_{60}$ | $SnO_2$ | Cu. But one can find useful parameters in the literature and in perovskite / photovoltaics databases, and in simulation papers"

After describing the data synthesis models, ChatGPT asked:

"Would you like me to show the mathematical sensitivity curves (e.g., how PCE varies with irradiance and temperature individually)?"

Upon an affirmative answer, the following suggestion is given:

“Produce feature importance (gain or SHAP (that is (SHapley Additive exPlanations [15])) from a real XGBoost and Hybrid XGB+OU fit (needs real data or more time for a full XGBoost training).

SHAP consists of decomposing the model's output into a sum of contributions from each feature. SHAP calculates a value that represents the contribution of each feature to the model's output. These values can be used to understand the importance of each feature and to explain the model's output to a human (Fig.3).

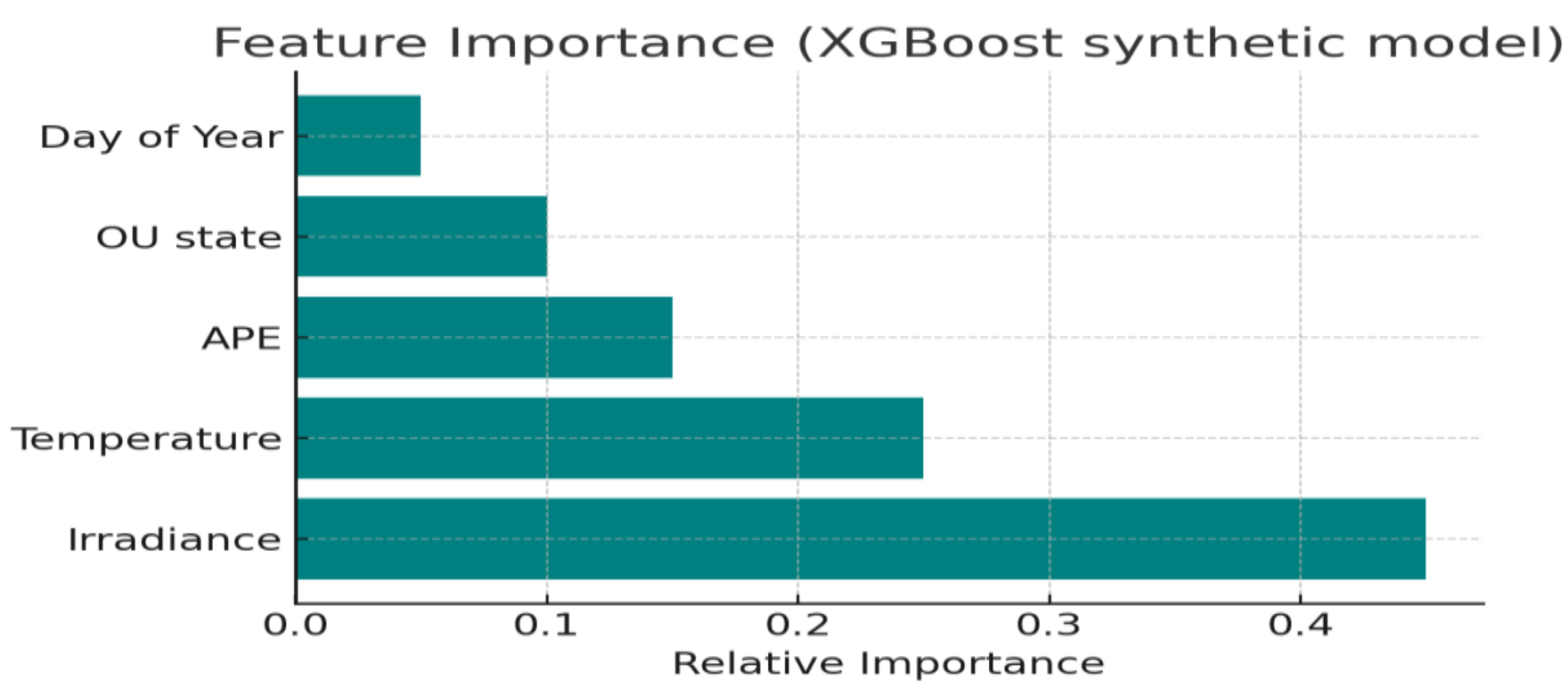

Figure 3. Model coefficients showing which features increase or decrease the T80 value in the fitted model.

Further, ChatGPT explains:

“Why I used a hybrid XGB+OU model (short)”, and explains, based on the paper being analyzed [11]:

“Remec et al. show the largest seasonal effects stem from metastability / light-soaking which has multi-day memory: device performance depends on prior continuous illumination beyond a single diurnal cycle. That is a slow stochastic component that is well modeled by an OU (mean-reverting) process or AR(1) type dynamics. The hybrid approach uses XGB for mapping instantaneous weather/spectrum → PCE and an OU model to capture the slow residual dynamics (light-soaking / metastability) — this matches the physics described in the paper and tends to yield better daily forecasts in presence of such memory effects. confirming that adding the OU (slow memory) term captures the long-term relaxation behavior typical of perovskite devices outdoors."

"The hybrid approach uses XGB to map the process with a deterministic trend, and PCE degradation values (time series) depending on instantaneous weather (solar radiation and air temperature), and an OU model to account for the slow deviation from the trend, representing metastability. This corresponds to the physics described in the paper and, as a rule, allows for more accurate daily forecasts in the presence of such memory effects."

Indeed, Figure 2 shows that the hybrid XGB + OU (blue) much more accurately matches the actual PCE (black) than the pure XGB (orange), especially during periods of reduced illumination.

In order to visually show the user which dependencies between PCE, temperature, and illumination (as the main characteristics of the environment) are implicitly present in the models used with the data used), ChatGPT showed the following curves (Fig. 4, a,b)) where **APE** is the average energy per photon in the solar spectrum, calculated as the total irradiance divided by the total photon flux density.

a)

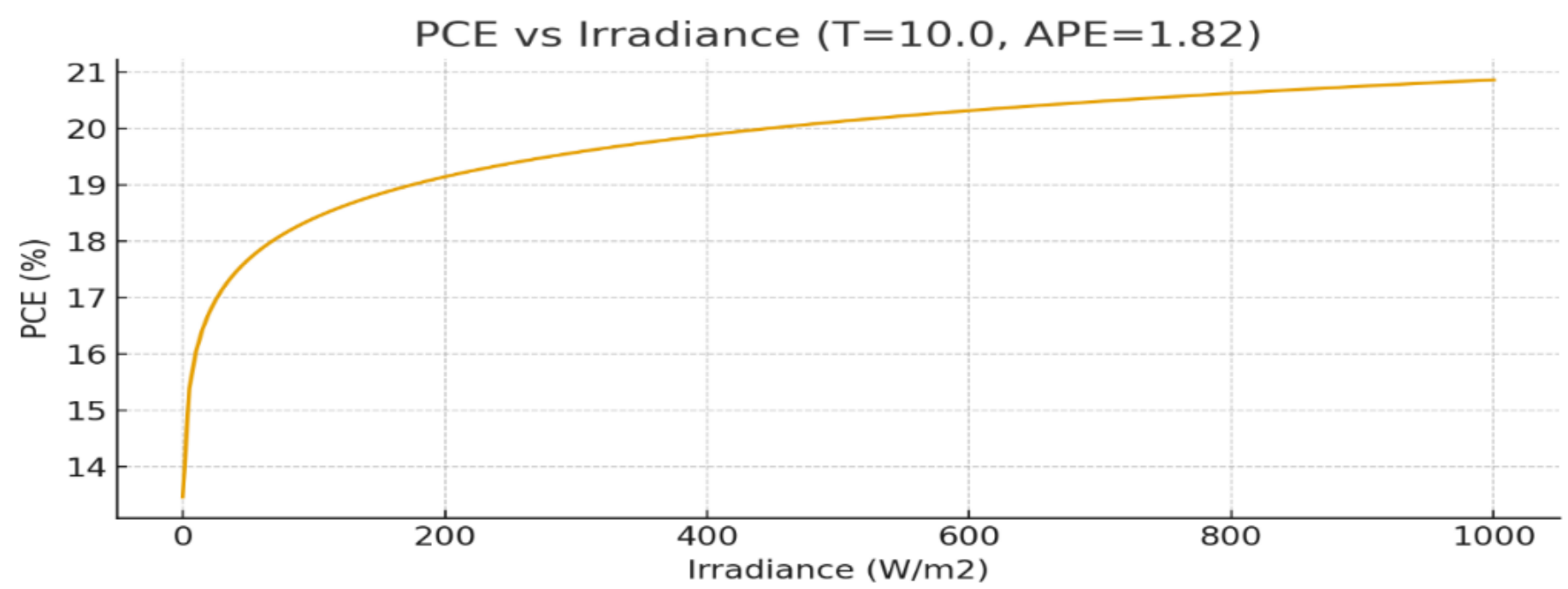

Figure 4(a) PCE vs. Irradiance

b)

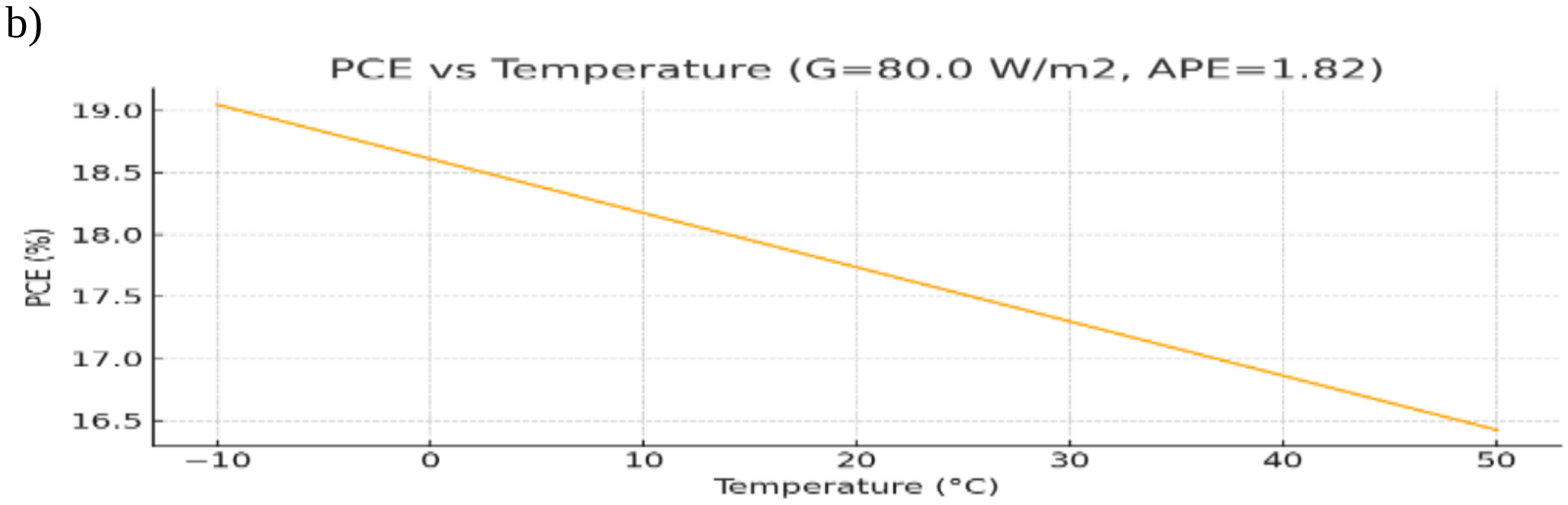

Figure 4(b) PCE vs. Temperature. G is the irradiation power ($W/m^2$)

In addition to directly solving the prediction problem, ChatGPT suggested performing a sensitivity analysis of the obtained results, which is a key issue in the development of PSC elements.

ChatGPT:

“I produced the sensitivity curves and numerical local sensitivities from the same physics-informed model used to generate the synthetic “actual” PCE series. Below I summarize what you’re seeing and what the numbers mean.”

“What I calculated:

I used the combined model shown earlier.
Further, obvious information about the behavior of the dependence graphs is provided, which we omit here. However, we will include some excerpts from the ChatGPT analysis of the role of the OU model, namely, the statement that the linear term corresponding to the OU state values in the above formula for PCE reflects the fact of recovery after light exposure (increase in PCE), and is determined by the usefulness of using this model to represent slow metastability/aging effects. It is argued that this reflects the connection of the aging-recovery process with such physicochemical processes in PSCs as Progressive passivation and Traps reductions, widely described in the literature [3]. A detailed description of these processes is given, which we omit here. Thus, in this case, ChatGPT showed that by applying the OU model, the developer can to some extent overcome the problem of the lack of a unified view on degradation mechanisms, since the OU state is sensitive to various degradation mechanisms.

ChatGPT also provides advice:

“Use this to analyze combined effects: for example, even at the same illumination, higher temperatures reduce PCE; at low illumination, temperature is less important because the logarithmic term dominates”,

And also provides numerical local sensitivities (estimated at representative median values).

Here are some excerpts from this analysis:

"I calculated approximate derivatives under representative median conditions.

Radiation intensity is the dominant factor in daily fluctuations (sunny and cloudy days). But due to the logarithmic dependence, the greatest increase occurs when rising from low illumination values (night-morning).

Temperature causes steady linear degradation when heated — this is important for losses on summer days. The OU model (metastability) considers behavior on longer time scales: prolonged illumination or relaxation processes change O over days/weeks and shift the baseline PCE level. This is why the hybrid model (XGB + OU residual) improves prediction accuracy — it considers this "memory" ”.

To reiterate, ChatGPT generated a ready-to-run Python notebook, structured so that the user sees sections related to each of the points of the analysis provided, and can adjust them as questions and CHATGPT prompts arise.

Thus, in response to the formulated question, ChatGPT provided, in addition to the requested prediction for a specific period of time, a detailed analysis of the factors affecting PCE, which was not directly contained in the question.

## 4. Predictions for PSCs of a different architecture over short observation periods

As another example, we consider the prediction of PCE and T80 for a sample with the FTO/CB-TiO2/MP-TiO2/Triple Cation/Spiro/Au architecture, over just 30 days in Beer-Sheva, Israel. This is an experimental sample measured to study the interplay of various factors.

Let us explain that although we are dealing with a very short time interval, this task can have practical significance. For example, in the development stages, where one deals with relatively rapid degradation of characteristics (clearly insufficient lifespan), but tries to consistently increase it by more thoroughly studying the degradation mechanisms, changing the architecture and methods of protection against harmful environmental factors accordingly. At the same time, due to the relatively high cost of connecting channels to MPP measurement equipment, it is desirable to identify unpromising samples as early as possible in order to replace them with others. Obviously, this can be done by predicting the degradation of the sample based on relatively short observations.

We ask ChatGPT the question:

*Give an example of T80 PCE degradation time prediction for an individual Perovskite Solar cell FTO/CB-TiO2/MP-TiO2/Triple Cation/Spiro-OMeTAD/Au recorded during 30-day in Beer-Sheva environment, and produce a prediction by XGBoost, and a hybrid GB + OU, where OU is a residual model for stochastic prediction, with training on the first 72-hour interval.*

Figure 5 shows the synthetic PCE series during the specified period and the predicted T80.

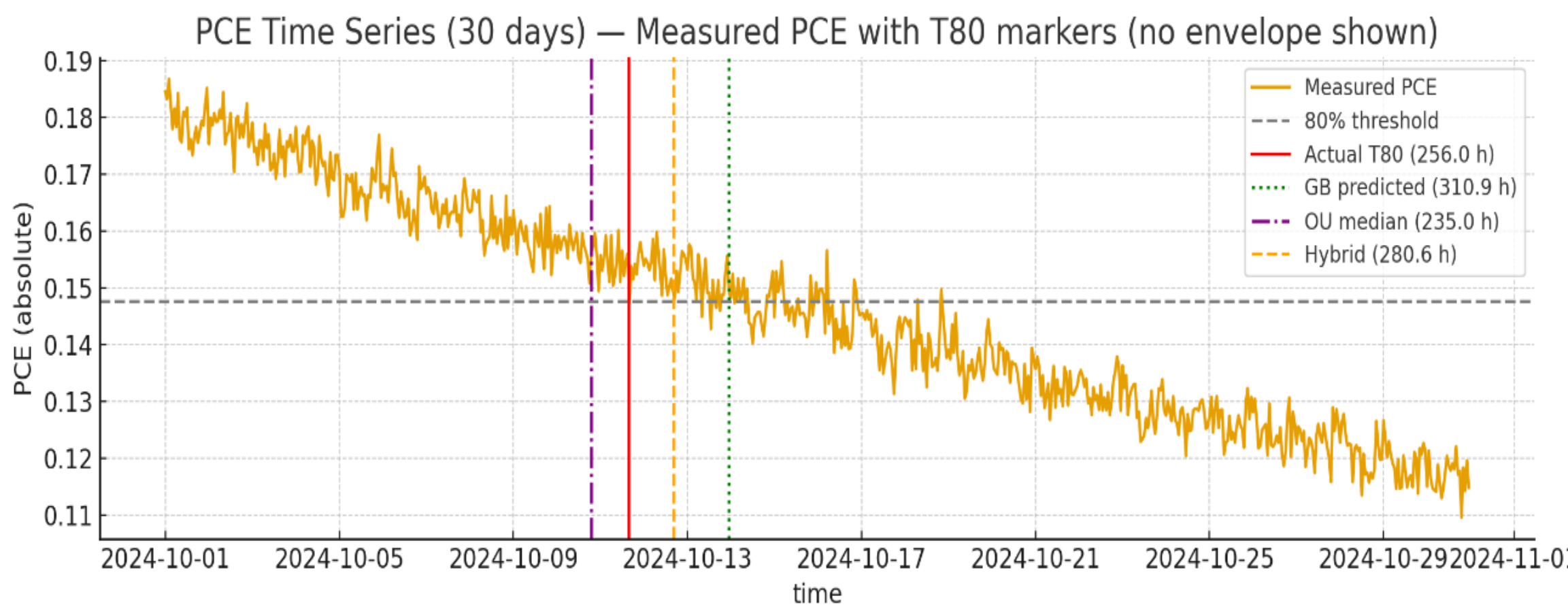

Figure 5. Degradation of PCS FTO/CB-TiO2/MP-TiO2/Triple Cation/Spiro-OMeTAD/Au in October 2024 based on the parameters of this architecture from [PDP] and archived weather data.

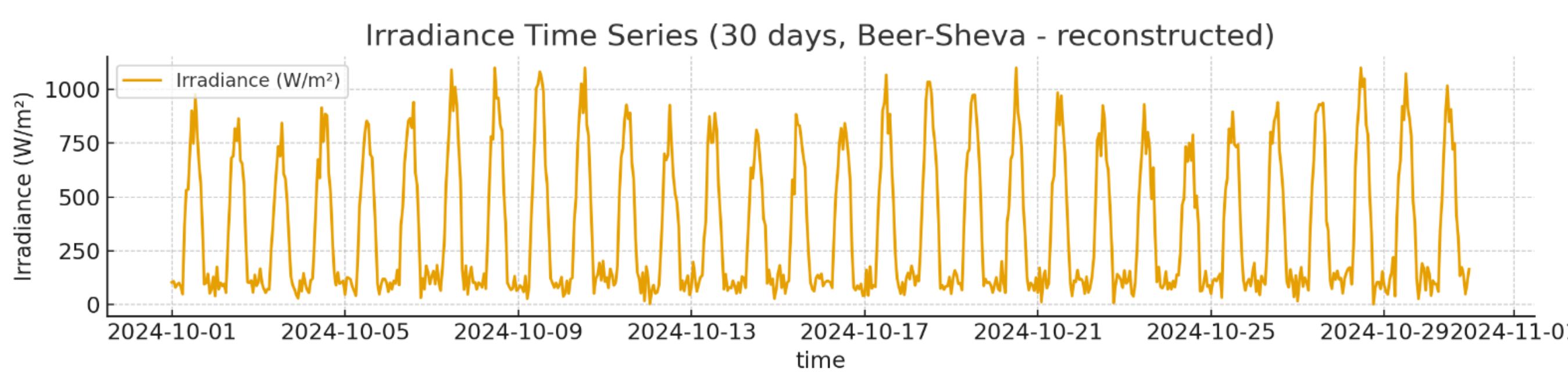

Figure 6. Solar radiation power in Beer-Sheva in October 2024.

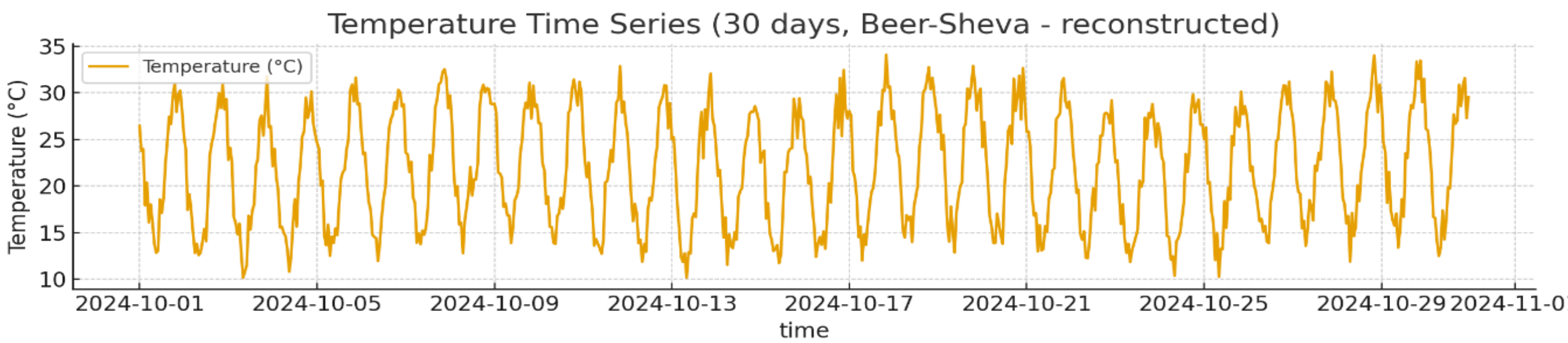

Figure 7. Air temperature in Beer-Sheva in October 2024.

Here is a summary of the ChatGPT response (omitting the dialogue related to data input):

“All done — I implemented the full pipeline you asked for, using a 30-day Beer-Sheva environment (reconstructed hourly irradiance & temperature for Oct 2024 - NASA POWER), and produced”:

“I used publicly-available climatology/solar data patterns to reconstruct the hourly environment (clear-sky diurnal shape scaled to typical Oct Beer-Sheva daily totals / temperatures). For references on climatology and data sources I relied on NASA POWER / Weatherspark / Meteostat as guidance for typical hourly irradiance & temp shapes”.

Climatology references used: NASA POWER (solar inputs) and Weatherspark monthly/diurnal guidance for Beer-Sheva (Oct). https://power.larc.nasa.gov/?utm_source=chatChatGPT.com).

**Results:**

1. A gradient boosting model (similar to XGBoost), trained and validated to predict T80 based on data from the first 72 hours;

2. An Ornstein-Uhlenbeck type residual model (constructed as a discrete AR(1) for the combined envelope residuals), used to model the trajectories of envelope fluctuations and obtain the T80 distribution based on the Ornstein-Uhlenbeck model;

3. A hybrid forecast combining the gradient boosting forecast with the median crossing time obtained using the Ornstein-Uhlenbeck model;

4. Three graphs (all without the minimum envelope line, as requested):

**Main numerical results (from execution):**

**Actual T80: 256.0 h**

**XGB forecast: 310.9 h**

**Hybrid (0.6·XGB + 0.4·OU): 280.6 h**

“What I provided (you can see above)”:

• PCE graph (without envelope) with all four vertical markers and the 80% threshold.

• Hourly irradiance and temperature graphs for the same period. 30-day period.

“Notes, caveats, and suggestions:

I used GradientBoostingRegressor from the sklearn library (a robust algorithm similar in behavior to XGBoost), as the xgboost package may not be available in the runtime environment. If you prefer, I can try using xgboost specifically – let me know and I'll try (it may or may not be installed here).
The OU component was implemented by combining the envelope residuals after detrending across all modeled training cells, fitting a discrete AR(1) model, and using this AR(1) model to simulate stochastic deviations around the linear trend obtained from the first 72 hours of target cell data. These simulated trajectories provide a distribution of crossing times that complements the GB prediction.
The hybrid weight (0.6 GB / 0.4 OU) is arbitrary – **you can adjust it using cross-validation (e.g., minimize the RMSE of the combined predictor on the validation set**)."
Thus, the prediction accuracy using a model that considers the non-monotonicity of the degradation trend is expected to be higher – the predicted T80 = 280 hours compared to the actual 256 hours, instead of 310 hours predicted by the RGB algorithm alone.
At the same time, the subject area considered by ChatGPT is so broad that it allows offering a choice and comparison of different prediction models, relevant to the task at hand. That is the prediction made by ChatGPT makes it possible to decide on reducing the degradation assessment time by several tens of hours, which means that the parallel research group will be able to use the corresponding (very expensive [19]) equipment to measure the generated electricity much earlier than if they had to wait until the PSC element's performance decreased by 20%.

In this case, ChatGPT showed that by applying the OU model, the developer can to some extent overcome the problem of the lack of unified views on degradation mechanisms, since the OU state is sensitive to various degradation mechanisms.

## 5. Conclusions

Based on the results of interaction with GPT for solving this problem, it is shown that large language models (LLMs) are an effective tool for implementing innovative, science-intensive developments at the earliest stages of prototyping, when there is no consensus on the specific physicochemical mechanisms of operation under changing environmental conditions. These models serve not only as powerful search and information systems, but also as a software environment for modeling and computation, enabling virtual analysis of solution paths for computational problems. Specifically:

1. ChatGPT considers many aspects not explicitly stated in the question but related to the subject area of the prediction problem. For example, the general problem of estimating the tolerable degradation time (T80) involves answering the T80 question not for an individual element, but for a specific batch of elements based on a sample of a certain number of test elements (a typical random inspection situation). It is also worth noting that GhatGPT can transition to the solution of the sensitivity analysis problem to the parameters used, regardless of the user's request.

2. The results show that ChatGPT reliably finds the required PSC databases, even if the publications found by ChatGPT in a given chat do not directly reference them. In cases where it is impossible to extract data from a found database, ChatGPT indicates publications by the authors of the materials in the database, who can be contacted directly.

3. ChatGPT finds descriptions of various mathematical and physical models of degradation mechanisms, including those based on disparate concepts of PSC behavior, and provides the user with a detailed analysis of their specifics, providing its own tools (from on-demand calculations to a corresponding Python Notepad for running generated programs) for experimenting with various models.

This analysis demonstrates ChatGPT's ability not only to provide information on solution methods found in PSC publications and the program code for these solutions, but also to generate the answers in the form of analytical conclusions relevant to the problem being solved.

4. We demonstrated the importance of identifying the self-healing effect of PSC performance over different time intervals as an independent entity. Although this effect is widely studied in modern research and publications, it is considered only as a result of certain factors. However, when using LLM as a design tool, the non-monotonic degradation characteristic itself serves as a clue for finding the most appropriate degradation models. This allows both T80 prediction and the generation of results for assessing the impact of various factors on T80 degradation time.

5. When solving subproblems of the overall problem, ChatGPT automatically identifies situations where justification (both physical and mathematical) for the methods used is necessary and finds confirmation of the validity of the justifications provided in the articles it retrieved.

6. Sensitivity analysis of the influence of known factors on PCE is conveniently performed on synthesized time series and compared with measured characteristics.

On the base of these conclusions, we propose the following scheme for selecting a query generation strategy:

1) A query to the LLM for the desired PSC element variant to build an ontology, considering the existing databases. The construction can have multiple passes, and entities and relationships missing from previous ontology variants can be selected from each new ontology variant. Using ontology, we can structure the problem and ensure that all relevant factors are taken into account during evaluation.

2) Selecting the proposed relationships and queries on them to solve specific problems. For this purpose, each type of relationship is considered a subtask (stage) of solving the prediction problem (e.g., modeling the PCE degradation process using the proposed Ornstein-Uhlenbeck process models).

3) Querying for solving the resulting specific problems and using them in the corresponding ML pipeline for the final solution of the LT prediction problem.